\documentclass[aps,showpacs,floatfix,prl,twocolumn]{revtex4}
\usepackage{amssymb,amsfonts,amsmath,bm}
\usepackage{graphicx}

\makeatletter

\renewcommand{\@makefntext}[1]{\parindent=1em\noindent\hbox to 1.8em{\hss$^{\@thefnmark}$}#1}
\renewcommand{\@footnotemark}{\hbox{\mathsurround=0pt$^{\@thefnmark}$}}

\makeatother

\begin{document}

\title{Angular momentum content of the $\rho$-meson in lattice QCD}

\author{Leonid Ya.\ Glozman, C.\! B.\ Lang, and Markus Limmer}
\affiliation{Institut f\"ur Physik, FB Theoretische Physik, 
Universit\"at Graz, A-8010 Graz, Austria}

\begin{abstract}
The variational method allows one to study the mixing of  interpolators with
different chiral transformation properties in the non-perturbatively determined 
physical state.  It is then possible to define and calculate in a
gauge-invariant manner the chiral as well as the partial wave content of the
quark-antiquark component of a meson in the infrared, where mass is generated.
Using a unitary  transformation from the chiral basis to the LSJ basis one may 
extract a partial wave content of a meson. We present results for the ground
state of the $\rho$-meson using quenched simulations as well as simulations with
$n_f=2$ dynamical quarks, all for lattice spacings close to 0.15 fm. We point
out that these results indicate a simple $^3S_1$-wave composition of the
$\rho$-meson in the infrared, like in the SU(6) flavor-spin quark model. 
\end{abstract}


\maketitle
%%%%%%%%%%%%%%%%%%%%%%%%%%%%%%%%%%%%%%%%%%%%%%%%%%%%%%%%%%%%%%%%%%%%%%%%%%%%%%%%

\paragraph{1. Motivation.}

The composition of hadronic states in quantum field theory is a subtle issue. 
Whereas in non-relativistic approaches the notion of a wave function and a
complete  basis of states is well-defined, in quantum field theory beyond the
ground state there is no well-defined single hadron, and the state is always a
scattering state with superposition of many particle components. A given hadron
interpolator couples in principle to all states with its quantum numbers. 

Lattice studies of hadrons are typically limited to spectroscopy, static
observables like magnetic moments, axial couplings, etc., as well as dynamical
observables such as form factors, parton distributions, etc. A challenging task
is to reveal the composition of hadrons, i.e., to understand the hadron structure in
ab-initio QCD calculations. Of course, in principle any hadron contains a large
amount of different Fock components and it is hardly possible to reconstruct on
the lattice the complete hadron wave function. Nevertheless, the analyses of the
phenomenological data and modeling of hadrons suggest that typically only a few
components are the dominant ones. To understand in ab-initio calculations
the structure of the leading component in the infrared, i.e., at the scale where
hadron mass is generated, would be an important step to improve the insight into hadron
structure.

There is a tool to study the hadron wave function on the lattice -- the
variational method \cite{Mi85,LuWo90} (for a recent, more complete set of
references see\ \cite{BlDeHi09}). The correlation function of a hadron
interpolator, which is built from a linear combination of many interpolating
operators with correct quantum numbers  will single out an optimal signal
associated with an exponential decay for large Euclidean time
distance. This combination defines the ``physical'' ground state hadron.   In
the variational method one chooses a set of interpolators $\{O_1,O_2,...,O_N\}$
that potentially couple to a given hadron and computes the correlation matrix
\begin{equation}
C(t)_{ij} = \langle O_i(t) O_j^\dagger(0) \rangle\;.
\end{equation}
If the set is complete enough the eigenvalues of a generalized eigenvalue
problem may be related to the eigenstates of the Hamiltonian. Then, if one is
interested in reconstructing the leading Fock component of a hadron in the
infrared, e.g., the $\bar q q$ component of a meson, one needs  interpolators
that allow one to define uniquely such a component. 

Let us consider the chiral limit of two-flavor mesons. All possible
quark-antiquark interpolators for non-exotic mesons can be classified according
to representations of the $SU(2)_L \times SU(2)_R$ and $U(1)_A$ groups
\cite{Gl03,Gl07}.  This basis is complete for a quark-antiquark system provided
that there is no explicit excitation of the gluonic field with non-vacuum
quantum numbers.  Using interpolators that belong to this basis allows one
to study the chiral symmetry breaking aspects in a meson. Diagonalizing the
cross-correlation matrix  one can reconstruct a decomposition of a given meson
in terms of different representations of  the chiral group. Chiral symmetry
breaking in the infrared would mean that the meson wave function has components
with different transformation properties with respect to $SU(2)_L \times
SU(2)_R$ and $U(1)_A$. 

It is also of interest to reconstruct a composition of a meson in terms of the
$^{2S+1}L_J$ basis, where ${\mathbf J} = {\mathbf L} + {\mathbf S}$ are the
standard
angular momenta. Such a decomposition provides a bridge to the language of the
quark model.   A priori it is clear that in the heavy quark limit the
non-relativistic language of the quark model is nearly adequate. There are well known achievements of the
quark model,  such as the $SU(6)$ flavor-spin symmetry, which is related to
rather large masses of the constituent quarks. These large masses are 
far from the tiny masses of the current quarks and it is
unclear what happens for light quarks. While there is a kind of
understanding that successes of the quark model are related to chiral symmetry
breaking and that these large masses of constituent quarks arise from the
coupling of current quarks to the quark condensate at low momenta, a
model-independent and gauge-invariant view on this problem is absent. Then, it
would be intriguing to see a decomposition of the leading quark-antiquark Fock
component of a meson in terms of the $^{2S+1}L_J$ basis in the infrared.

There is a possibility to establish the angular momentum decomposition of the
leading quark-antiquark component. Both the chiral basis and the $^{2S+1}L_J$
basis are complete for a two-particle system. There exists a unitary
transformation from the chiral basis to the standard  $^{2S+1}L_J$ basis in the
center-of-momentum system \cite{GlNe07}. Each of the states in the relativistic
chiral basis can be uniquely represented in terms of the allowed $^{2S+1}L_J$
states. Diagonalizing the cross-correlation matrix in terms of the interpolators
with carefully chosen chiral transformation properties, and using this unitary
transformation to the $^{2S+1}L_J$ basis one can reconstruct a partial wave
decomposition of the leading Fock component of a meson.

The method is general and can be applied to any meson. Here we use
 as an example the $\rho$-meson. If chiral symmetry is unbroken, then
there are two possible states in the chiral basis with the $\rho$-meson quantum
numbers, $|(0,1)+(1,0);1 ~ 1^{--}\rangle$ and  $|(1/2,1/2)_b;1 ~ 1^{--}\rangle$,
where $(0,1)+(1,0)$ and $(1/2,1/2)_b$ specify two different representations of
$SU(2)_L \times SU(2)_R$ that are compatible with the $\rho$-meson quantum
numbers $I, J^{PC} = 1, 1^{--}$  (a subscript $b$  specifies
one of the two different representations $(1/2,1/2)_a$ and $(1/2,1/2)_b$ \cite{Gl03,Gl07}).
These two representations are a complete
set for particles with the quantum numbers of the $\rho$-meson. A unitary
transformation from the chiral basis to the  $^{2S+1}L_J$ basis takes the form 
\begin{eqnarray}\label{chiral_basis}
|(0,1)+(1,0);1 ~ 1^{--}\rangle&=&\sqrt{\frac{2}{3}}|1;
{}^3S_1\rangle+\sqrt{\frac{1}{3}}|1;{}^3D_1\rangle\;,\nonumber\\
 |(1/2,1/2)_b;1 ~ 1^{--}\rangle&=&\sqrt{\frac{1}{3}}|1;
{}^3S_1\rangle-\sqrt{\frac{2}{3}}|1;{}^3D_1\rangle\;.\nonumber\\
\end{eqnarray}

Hence, if we choose two different interpolators with the same spatial content
(i.e., with same smearing width of the quarks; its size defines the scale where 
we probe the hadron)
and transformation properties according to   $|(0,1)+(1,0);1 ~ 1^{--}\rangle$
and $|(1/2,1/2)_b;1 ~ 1^{--}\rangle$, we will be able to reconstruct the angular
momentum content of the $\bar q q$ component. Such interpolators are well-known
-- the quark bilinears $\bar{q}\gamma^i{\bm\tau}q$ and
$\bar{q}\sigma^{0i}{\bm\tau}q$
\cite{CoJi97}. It has been established in lattice simulations
that the $\rho$-meson couples to both (see, e.g., 
\cite{BeLuMe03,BrBuGa03,BuGaGl06} for quenched and  \cite{AlAnAo08} for dynamical
simulations). A key
property is that these two interpolators have radically different chiral
transformation properties. The former one transforms as $(0,\,1) + (1,\,0)$
while the latter one belongs to the $(1/2,\, 1/2)_b$  representation of $SU(2)_L
\times SU(2)_R$.  Consequently, by diagonalizing  the cross-correlation matrix
with the $\bar{q}\gamma^i{\bm\tau}q$ and $\bar{q}\sigma^{0i}{\bm\tau}q$
interpolators we can reconstruct a decomposition of a given eigenstate (i.e., of
the ground state $\rho$-meson and its excitations) in terms of the possible
partial waves in the infrared.

\paragraph{2. Analysis of lattice correlators.}

For completeness we briefly summarize the features of the variational analysis
that allow the decomposition of the  ground state properties in the
$\rho$-channel.

The normalized physical states $|n\rangle$ propagate in time with
\begin{equation}
\langle n(t)|m(0)\rangle =\delta_{nm}\mathrm{e}^{-E^{(n)} t}\:.
\end{equation}
The interpolating (lattice) operators $O_i(t)$ are projected to vanishing
spatial momentum and are usually not normalized. We compute the correlation function
\begin{equation}\label{corr_inf}
C(t)_{ij}=\langle O_i(t)O_j^\dagger(0)\rangle=\sum_n a_i^{(n)} a_j^{(n)*} 
\mathrm{e}^{-E^{(n)} t}\;,
\end{equation}
with the coefficients giving the overlap of the lattice operator with the
physical state,
\begin{equation}
a_i^{(n)}=\langle 0| O_i|n\rangle\;.
\end{equation}
For interpolating operators $O_i$ spanning an orthogonal basis these values
would indeed constitute the wave function of  state $|n(0)\rangle$ in that
basis.

We assume that the correlation matrix (\ref{corr_inf}) can be approximated by a
finite sum over  $N$ states and denote this  approximation by $\widehat
C(t)_{ij}$.

It can be shown \cite{Mi85,LuWo90} (for a recent discussion see 
\cite{BuHaLa08,BlDeHi09}) that the generalized eigenvalue problem
\begin{equation}\label{gev_1}
\widehat C(t)_{ij} u_j^{(n)} =\lambda^{(n)}(t,t_0)\widehat C(t_0)_{ij} u_j^{(n)}
\end{equation}
allows to recover the correct eigenvalues and eigenvector within some approximation.
One finds
\begin{equation}\label{gev_2}
\lambda^{(n)}(t,t_0)=\mathrm{e}^{-E^{(n)} (t-t_0)} 
\left(1+\mathcal{O}\left(\mathrm{e}^{-\Delta E^{(n)} (t-t_0)}\right)\right)
\;,
\end{equation}
where $\Delta E^{(n)}$ may be as small as the distance to the next nearby energy
level. In \cite{BlDeHi09} it was pointed out that in an interval $t_0\le t \le
2\,t_0$ these contributions are suppressed and leading terms even have $\Delta
E^{(n)}$ equal to the distance to the first neglected energy level  $E_{N+1}$.
At $t_0$ all eigenvalues  are 1 and the eigenvectors are arbitrary.

The eigenvectors come out orthogonal (dual) to the original wave  functions
$a^{(n)}$,
\begin{equation}\label{gev_dual}
(u^{(n)},a^{(m)})\equiv\sum_{i=1}^N u_i^{(n)*} a_i^{(m)}= c^{(m)}\delta_{nm}\;,
\end{equation}
and approximate the correct ones. Here $c^{(m)}$ is a normalization which we
get rid off below.

We define a sum of lattice operators
\begin{equation}
\eta^{(n)} \equiv \sum_{i=1}^N u_i^{(n)*} O_i\;,
\end{equation}
and find
\begin{eqnarray}
\langle 0|\eta^{(n)}|m\rangle&=&\sum_{i=1}^N u_i^{(n)*} \langle 0|O_i|m\rangle\nonumber\\
&=&\sum_{i=1}^N u_i^{(n)*} a_i^{(m)}=c^{(n)}\delta_{nm}\;.
\end{eqnarray}
Therefore
\begin{equation} 
\eta^{(n)\dagger}|0\rangle=c^{(n)*}|n\rangle\;.
\end{equation}
The eigenvector coefficients are related to the  composition of the eigenstate
in terms of the interpolating operators. The original values $a_i^{(n)}$ are
then
\begin{eqnarray}
a_i^{(n)}=\langle 0|O_i|n\rangle
&=&\frac{1}{c^{(n)*}}\langle 0|O_i\,\eta^{(n)\dagger}|0\rangle
\nonumber\\
&=&
\frac{1}{c^{(n)*}}\sum_{j=1}^N u_j^{(n)} \langle 0| O_i O_j^\dagger|0\rangle\;.
\end{eqnarray}
They would agree with $u_j^{(n)}$ if the interpolators were orthogonal, which
they are usually not.

However, with (\ref{corr_inf}) and (\ref{gev_dual})
we find for the large $t$ behavior (summation convention)
\begin{eqnarray}
w_i^{(n)}\equiv \widehat C(t)_{ij} u_j^{(n)}
&=&c^{(n)*} a_i^{(n)} \mathrm{e}^{-E^{(n)} t}\;,\nonumber\\
\left(u^{(n)},w^{(n)}\right)=
u_i^{(n)*} \widehat C(t)_{ij} u_j^{(n)}&=& c^{(n)*} u_i^{(n)*} a_i^{(n)}
\mathrm{e}^{-E^{(n)} t}\nonumber\\
&=&\left|c^{(n)}\right|^2 \mathrm{e}^{-E^{(n)} t}\;,
\end{eqnarray}
and for the ratio for large $t$
\begin{equation}
\frac{\left|w_i^{(n)}\right|^2}{(u^{(n)},w^{(n)})}
= \left|a_i^{(n)}\right|^2 \mathrm{e}^{-E^{(n)}t}\;.
\end{equation}
Assuming asymptotically leading exponential behavior this allows to read off
$|a_i^{(n)}|$ in the asymptotic region.

The ratio
\begin{equation}
\left|\frac{a_i^{(n)}}{a_i^{(m)}}\right|^2=
\left|\frac{\langle 0|O_i(t)|n\rangle}{\langle 0|O_i(t)|m\rangle}\right|^2
\end{equation}
tells us how much the interpolating operator $O_i$ contributes to the
eigenstates $|n\rangle$ and $|m\rangle$. This can be used to discuss the ratio
of decay constants of various excitations as done in 
\cite{BuEh07,BuHaLa08,BlDeHi09}.

If we are interested in ratios of couplings of the different lattice
operators to the physical states, we may utilize
\begin{equation}\label{ratio_op_comp}
\frac{C(t)_{ij} u_j^{(n)}}{C(t)_{kj} u_j^{(n)}}=\frac{a_i^{(n)}}{a_k^{(n)}}\;.
\end{equation}
This ratio tells us how much different interpolating operators contribute to the
eigenstate $|n\rangle$. This can be used to discuss contributions of, e.g.,
different representations of the vector meson channel.

\paragraph{3. Lattice simulation and results.}

In a series of papers we have studied the hadron spectrum derived in the
quenched case \cite{BuGaGl04a,BuGaGl06,GaGlLa08} as well as for dynamical
fermions  \cite{GaHaLa08}. The gauge field action was the L\"uscher-Weisz action
and the fermions were simulated with the so-called chirally improved (CI) Dirac
operator \cite{Ga01a,GaHiLa00}.

In these analyses the variational method was used; the hadron interpolators were
built from smeared quark sources. Here we refer only to a subset of these
results, namely the $\rho$-channel ($J^{PC}=1^{--}$) and restrict ourselves to
the four interpolators
\begin{eqnarray}
&O_1=\overline u_n \gamma^i d_n\;,\;\;
&O_2=\overline u_w \gamma^i d_w\;,\;\;\\
&O_3=\overline u_n \gamma^t \gamma^i  d_n\;,\;\;
&O_4=\overline u_w \gamma^t \gamma^i  d_w\;.
\end{eqnarray}
Here $\gamma^i$ is one of the spatial Dirac matrices, $\gamma_t$ is the
$\gamma$-matrix in (Euclidean) time direction, and the subscripts $n$ and $w$
(for narrow and wide)  denote the two smearing widths, 0.25 fm and 0.41 fm,
respectively \cite{GaGlLa08}.

\begin{table}[tb]
\caption{\label{tab_data}
Specification of the data used here; for the gauge coupling only the leading
value $\beta_{LW}$ is given, $m_0$ denotes the bare mass parameter of the
CI-action. Further details like  the determination of the lattice spacing and
the $\pi$- and $\rho$-masses are found in \cite{GaGlLa08,GaHaLa08}. For the
quenched case and ensemble A we used 100 configurations, for sets B and C we
analyzed 200 configurations each. The lattice size is $16^3\times 32$.}
\begin{center}
\begin{tabular}{lccccc}
\hline
\hline
Data&  $\beta_{LW}$ &  $a\,m_0$& $a$ [fm] & $m_\pi$[MeV]& $m_\rho$[MeV]\\
\hline
Quenched& 7.90 & 0.04--0.20&   0.1480(10)   &475--1053     &  912--1251\\
dyn.: A & 4.70 & -0.050&       0.1507(17) &526(7)& 922(17)\\
dyn.: B & 4.65 & -0.060&       0.1500(12) &469(4)& 897(13)\\
dyn.: C & 4.58 & -0.077&       0.1440(12) &318(5)& 810(28)\\
\hline
\hline
\end{tabular}
\end{center}
\end{table}

All results have been obtained on lattices of size  $16^3\times 32$ with lattice
spacing $a$ close to 0.15 fm (see Table \ref{tab_data}). The runs A, B and C are
for $n_f=2$ mass degenerate CI-fermions.

\begin{figure}[t]
\begin{center}
\includegraphics*[width=7cm]{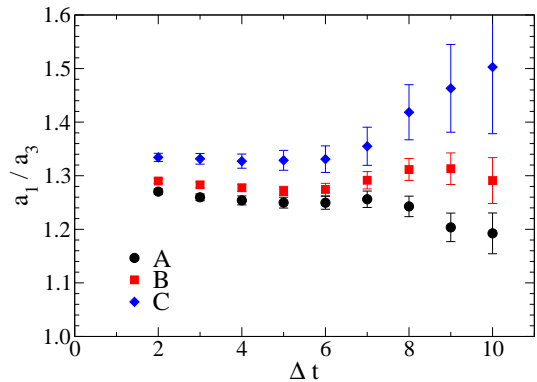}
\end{center}
\caption{\label{fig_plateaus}
As an example we show the ratio of the contributions of $O_1$ vs. $O_3$ (the
narrow-narrow smeared operators) for ensembles A, B and C.  All four operators
have been included in the correlation matrix. We do not show the points for
$\Delta t\le t_0=1$ and above $\Delta t=10$, where approaching the time-symmetry
points the results become statistically unstable. The error bars have been
determined with single-elimination jack-knife.}
\end{figure}

\begin{figure}[th]
\begin{center}
\includegraphics*[width=7cm]{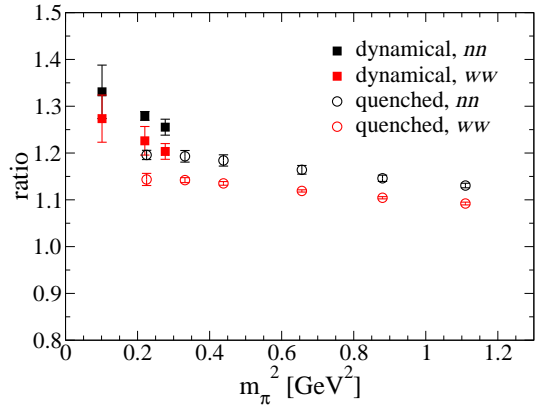}
\end{center}
\caption{\label{fig_ratio} We show the plateau mean values of the ratios 
$a_1/a_3$ and $a_2/a_4$ for the  quenched (circles) and dynamical (filled
squares) data sets considered in Table \ref{tab_data}. For the dynamical data we
also have partially quenched results, i.e., with $m_{valence}>m_{sea}$ (not
shown here), which approach the purely quenched values for pion masses above
$\approx$0.7 GeV. The error bars have been determined with single-elimination 
jack-knife.}
\end{figure}

As detailed in \cite{GaGlLa08,GaHaLa08} it is possible to identify the ground
state and the first excited state. In that study we also presented  the
corresponding eigenvector components ($u_j^{(1)}$ in our present notation) which
were stable over several time slices. Continuing that analyses we now utilize
Eq.\ (\ref{ratio_op_comp})  and determine the ratios ${a_i}/{a_k}$ for the ground
state (we have dropped the superscript reference to the physical state), in
particular  ${a_1}/{a_3}$ and ${a_2}/{a_4}$ which give information on the
relative contributions of operators with $\gamma^i$ compared to
$\gamma^t\gamma^i$. 

We chose the relatively small value $t_0=1$. Larger values lead to  larger
errors; however, we found very nice plateau behavior over a  range of
propagation distance in Euclidean time $t$. In Fig.\ \ref{fig_plateaus} we show
the surprising stability of the plateau of the ratio ${a_1}/{a_3}$ for the three
data sets obtained for dynamical fermions. We average the values in the range $3
\le \Delta t \le 7$ and show the resulting ratio in Fig. \ref{fig_ratio} for
both, the quenched and the dynamical data. We find different behavior towards
the chiral limit, indicating the effect of the fermion sea. 

Extrapolation towards the physical point indicates a value close to 1.35,
i.e., compatible with $\sqrt{2}$ within
the errors. Consequently chiral symmetry is broken
such that the $\bar q q$ component of  the $\rho$-meson is a superposition of
the $(0,1)+(1,0)$ and the $(\frac{1}{2},\frac{1}{2})_b$ representations with the
ratio close to $\sqrt{2}$. Inverting the unitary matrix in (\ref{chiral_basis})
we then conclude that the physical $\rho$-meson in the infrared is almost purely
a $^3S_1$ state. This gives a model-independent and gauge invariant explanation
of the success of the SU(6) flavor-spin symmetry for the $\rho$-meson. 

This result holds in the infrared where the mass is generated. We probe the chiral 
and partial wave content of the $\rho$ wave function at a scale fixed by the 
smearing size. At higher resolution ($Q^2\to\infty$) the $\beta$-function approaches 
zero, the pseudotensor current decouples from the $\rho$-meson and
the partial wave decomposition should be determined by the $(0,1)+(1,0)$ representation.
Such a tendency is indicated in Fig.~\ref{fig_ratio}. It is also consistent
with the ratio $\approx1.6$ obtained in dynamical calculations with
point interpolators (i.e., a scale fixed by $a=0.114$ fm) in \cite{AlAnAo08}.

\begin{acknowledgments}
We thank G.\ Engel, C.\ Gattringer and D.\ Mohler for discussions. L.Ya.G.\! and M.L.
acknowledge support of the Fonds zur F\"orderung der Wissenschaflichen Forschung
(P19168-N16) and (DK W1203-N08), respectively. The calculations have been performed on the SGI
Altix 4700 of the Leibniz-Rechenzentrum Munich and on
local clusters at ZID at the University of Graz.
\end{acknowledgments}


\vspace*{-5mm}

\begin{thebibliography}{10}

\bibitem{Mi85}
C.~Michael,
\newblock Nucl.\ Phys.\ B {\bf 259}, 58 (1985).

\bibitem{LuWo90}
M.~L{\"u}scher and U.~Wolff,
\newblock Nucl.\ Phys.\ B {\bf 339}, 222 (1990).
%%CITATION = NUPHA,B339,222;%%

\bibitem{BlDeHi09}
B.\ Blossier, M.~DellaMorte, G.~von Hippel, T.~Mendes, and R.~Sommer,
\newblock JHEP {\bf 0904}, 094 (2009), arXiv:0902.1265 [hep-lat].
%%CITATION = 0902.1265;%%

\bibitem{Gl03}
L.~Y.\ Glozman,
\newblock Phys.\ Lett.\ B {\bf 587}, 69 (2004), hep-ph/0312354.
%%CITATION = HEP-PH/0312354;%%

\bibitem{Gl07}
L.~Y.\ Glozman,
\newblock Phys.\ Rep.\ {\bf 444}, 1 (2007), hep-ph/0701081.
%%CITATION = HEP-PH/0701081;%%

\bibitem{GlNe07}
L.~Y.\ Glozman and A.~V.\ Nefediev,
\newblock Phys.\ Rev.\ D {\bf 76}, 096004 (2007), arXiv:0704.2673 [hep-ph].
%%CITATION = 0704.2673;%%

\bibitem{CoJi97} 
T.~D.\ Cohen and X.\ Ji, 
\newblock  Phys.\ Rev.\ D {\bf 55}, 6870 (1997), hep-ph/9612302.
%%CITATION = hep-ph 9612302;%%

\bibitem{BeLuMe03}
D.~Becirevic, V.~Lubicz, F.~ Mescia, and C.~Tarantino,
\newblock JHEP {\bf 0305} 007 (2003), hep-lat/0301020.
%%CITATION = hep-lat 0301020;%%

\bibitem{BrBuGa03}
V.~M.~Braun et~al.,
\newblock Phys.\ Rev.\ D {\bf 68}, 054501 (2003), hep-lat/0306006.
%%CITATION = hep-lat 0306006;%%

\bibitem{BuGaGl06}
T.~Burch et.~al.,
\newblock Phys.\ Rev.\ D {\bf 73}, 094505 (2006), hep-lat/0601026.
%%CITATION = hep-lat 0601026;%%

\bibitem{AlAnAo08}
C.~Allton et~al.,
\newblock Phys.\ Rev.\ D {\bf 78}, 114509 (2008), arXiv:0804.0473 [hep-lat].
%%CITATION = 0804.0473;%%

\bibitem{BuHaLa08}
T.~Burch, C.~Hagen, C.~B.\ Lang, M.~Limmer, and A.~Sch{\"a}fer,
\newblock Phys.\ Rev.\ D {\bf 79}, 014504 (2009), arXiv:0809.1103 [hep-lat].

\bibitem{BuEh07}
T.~Burch and C.~Ehmann, Nucl. Phys. A {\bf 797}, 33 (2007), hep-lat/0701001.

\bibitem{BuGaGl04a}
T.~Burch et~al.,
\newblock Phys.\ Rev.\ D {\bf 70}, 054502 (2004), hep-lat/0405006.
%%CITATION = HEP-LAT 0405006;%%

\bibitem{GaGlLa08}
C.~Gattringer, L.~Y.\ Glozman, C.~B.\ Lang, D.~Mohler, and S.~Prelovsek,
\newblock Phys.\ Rev.\ D {\bf 78}, 034501 (2008), arXiv:0802.2020 [hep-lat].
%%CITATION = 0802.2020;%%

\bibitem{GaHaLa08}
C.~Gattringer et~al.,
\newblock Phys.\ Rev.\ D {\bf 79}, 054501 (2009), arXiv:0812.1681 [hep-lat].
%%CITATION = 0812.1681;%%

\bibitem{Ga01a}
C.~Gattringer,
\newblock Phys.\ Rev.\ D {\bf 63}, 114501 (2001), hep-lat/0003005.
%%CITATION = hep-lat 0003005;%%

\bibitem{GaHiLa00}
C.~Gattringer, I.~Hip, and C.~B.\ Lang,
\newblock Nucl.\ Phys.\ B {\bf 597}, 451 (2001), hep-lat/0007042.
%%CITATION = HEP-LAT 0007042;%%

\end{thebibliography}
\end{document}